\documentclass[aps,prb,twocolumn,floatfix,floats,superscriptaddress,showpacs]{revtex4}
\usepackage{graphicx,latexsym}
\usepackage{amsmath}

\begin{document}

\title{Spectral properties and magneto-optical excitations in semiconductor double-rings under Rashba
spin-orbit interaction}

\author{Wen-Hsuan Kuan}
\affiliation{ Physics Division, National Center for Theoretical
Sciences, Hsinchu 30013, Taiwan, R.O.C.}
\author{Chi-Shung Tang}
\affiliation{ Research Center for Applied Sciences, Academia
Sinica, Taipei 11529, Taiwan, R.O.C.}
\author{Cheng-Hung Chang}
\affiliation{ Physics Division, National Center for Theoretical
Sciences, Hsinchu 30013, Taiwan, R.O.C.}
\affiliation{ Institute
of Phyics, National Chiao Tung University, Hsinchu 30013, Taiwan,
R.O.C.}


\begin{abstract}

We have numerically solved the Hamiltonian of an electron in a
semiconductor double ring subjected to the magnetic flux and Rashba
spin-orbit interaction. It is found that the Aharonov-Bohm energy
spectrum reveals multi-zigzag periodic structures. The
investigations of spin-dependent electron dynamics via Rabi
oscillations in two-level and three-level systems demonstrate the
possibility of manipulating quantum states. Our results show that
the optimal control of photon-assisted inter-ring transitions can be
achieved by employing cascade-type and $\Lambda$-type transition
mechanisms. Under chirped pulse impulsions, a robust and complete
transfer of an electron to the final state is shown to coincide with
the estimation of the Landau-Zener formula.

\end{abstract}

\pacs{74.25.Gz, 71.70.Ej, 74.25.Nf}
%74.25.Gz Optical properties
%71.70.Ej Spin-orbit coupling, Zeeman and Stark splitting, Jahn-Teller effect
%74.25.Nf Response to electromagnetic fields (nuclear magnetic resonance, surface impedance, etc.)
\maketitle

\section{Introduction}

The progress in epitaxial growth promotes the use of
low-dimensional semiconductor nanostructures in optoelectronic
devices. Investigations on fundamental physical properties such as
the electronic structure and the carrier population can be
directly measured and estimated from the photoluminescence
spectrum. Theoretically, several proposals on magneto-optical
studies for quantum dot systems have been put forward in the last
decade.\cite{dotwork} Nowadays, coherent optical manipulations of
single quantum systems have attracted further attention. The
mature technologies in optical control and measurements provide
the great opportunity to realize the quantum qubits as logical
gates\cite{logicalgate} on storage and quantum information
processing.\cite{Rabiqubit}

In 1990s, the progress of technology has enabled the experimental
study of a mesoscopic ring threaded by a static magnetic field
displays persistent currents,\cite{Levy90,Chan91} which oscillate as
a function of magnetic flux $\Phi$ with a period $\Phi_0 = hc/e$.
Recently, applications on spin-orbit interaction (SOI) originated
from the breaking of inversion symmetry that gives rise to intrinsic
spin splitting in semiconductor systems open a field of spintronics.
It was pointed out that the quantum transport of electrons in a
spin-polarized system differs greatly from that in a spin-degenerate
device.\cite{RMP76323} The utilization of the spin degree of freedom
offers the mechanism to speed up quantum information processing. In
nature electronic devices such as the Datta-Das
transistor,\cite{DattaDas} spin-waveguide\cite{spinWG} and
spin-filter\cite{spinF} were proposed.

Several theoretical works associated with Rashba SOI due to
structural inversion asymmetry in quantum dot systems were
studied.\cite{dotRashba}  More recently, the success on
self-assembled formation of concentric quantum double
rings\cite{NanoLetter} provides a new system to explore electron
dynamics by magneto-optical excitations on the basis of fully
analyzed signature of AB spectrum within the effect of Rashba SOI.
The radius of flat double rings is about 100\,nm with thickness
approximately 3\,nm. Therefore carriers are coherent all
throughout these small geometries. Within a time scale shorter
than the dephasing time,\cite{Eberly} the Rabi oscillation (RO)
can provide a direct control of excited state population
especially in strong excitation regime. It was proposed to be a
good optical implement in quantum dot systems.\cite{RO} However, a
simple two-level system involving Rabi oscillations with Rashba
SOI in a coaxial double quantum ring has not yet been studied.
Therefore, in this paper, we consider two-level and three-level
models to explore spin-dependent electron dynamics assisted by RO
processes. Under influences of the magnetic flux, the spin feature
of the system is demonstrated only through effects of Rashba SOI.
The presence of Rashba SOI also plays an important role in the
mixture of neighboring angular momenta as well as spins that
builds up a new selection rule and opens more dipole-allowed
transition routes.

In view of quantum algorithm realization, the two-level Rabi
oscillators are often the prototype of the qubit generators.
However, it is also important to establish coherent control in
realistic multi-level quantum
systems.\cite{chemical,smallmolecule}
%Applications, for example, on standard
%control approach for modelling molecular bonds in Morse and
%harmonic oscillators are still limited.
Hence, we explore the multi-level dynamical system involving the
cascade-type and the $\Lambda$-type three-level schemes driven by
either sinusoidal impulsions or chirped laser pulses.
%We focus on
%accelerating \textit{direct} inter-ring transitions and
%\textit{indirect} quantum tunneling via photon-assisted processes.
To achieve efficient transfers, we employ adiabatic rapid passage
method (ARP),\cite{ARP,pulse} namely that the excitation process
rapid compared with the natural life time of an excited state in
the limit of slowly varying detuning field, to simulate complete
transfer processes. We will show that the probability of an
electron that occupies the final state coincides with the
estimation of the Landau-Zener formula in the adiabatic
limit.\cite{LZ}

The paper is organized as follows: In Sec. II the single-particle
Hamiltonian is derived and numerically solved and SOI accompanied
AB energy spectrum is analyzed. In Sec. III ROs between two levels
selected from the double-ring spectrum are studied. In Sec. IV
photon-assisted electron transitions in the three-level systems of
the cascade- and $\Lambda$-schemes are investigated. Finally the
paper ends with a conclusion in Sec. V.

\section{The energy spectra of the double-ring system}

\begin{figure}[t!]
\includegraphics[width=0.40\textwidth,angle=0]{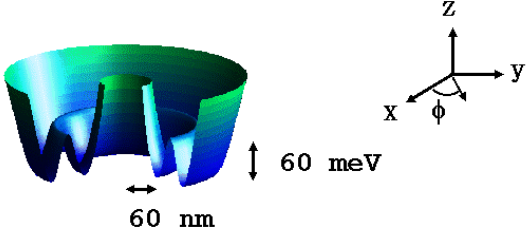}
\caption{\label{TotPot} (Color online) The diagram of part of the
double-ring potential depicted in $0 \leq \phi \leq 3\pi/2$. The
radius of the ring is about 160 nm and effective range of the
magnetic flux $r_{\Phi}$ is about 7 nm. }
\end{figure}
The system of a double-ring 2DEG is enclosed by a magnetic flux in
the presence of Rashba SOI. The electron is confined in an
axial-symmetric potential $V_c$, shown in Fig.~(\ref{TotPot}),
which is exposed in a monochromatic electromagnetic (EM) field. In
semiclassical description the Hamiltonian is given by
\begin{eqnarray}
 H &=&
 \frac{1}{2m^*}\left[\vec{p}- \frac{e}{c}{\mathbf{A}}(\vec{r}, t)\right]^2
 + V_c({r}) \nonumber\\
 &&+\frac{\alpha}{\hbar}[\vec{\sigma}\times(\vec{p}- \frac{e}{c}{\mathbf{A}}(\vec{r}, t))]_z
\end{eqnarray}
%\end{widetext}
where ${\mathbf{A}}(\vec{r}, t)$ contains contributions from the
magnetic flux and the EM field, and $\alpha$ is the coupling
constant of Rashba SOI.  The double-ring potential is modeled as
\begin{equation}
V_c(r) = \frac{1}{2}m^*\omega_0^2 (r-r_0)^2 + \sum_{i=1}^3 V_i
e^{-(r-r_i)^2/\sigma_i^2},
\end{equation}
where $\omega_0$ is a factor defining the characteristic length
$l_c=\sqrt{\hbar/m^*\omega_0}$ of the system and $\sigma_i$ is the
Gaussian spatial width. The magnetic flux applied through the
central region of the inner ring within $r_{\Phi}$ is described by
the vector potential $\vec{A}_{\Phi} = {Br}/2 \,\hat{\phi}$  for $r
\leq r_{\Phi}$ and $\vec{A}_{\Phi} = Br^2_{\Phi}/{2r} \,\hat{\phi}$
for  $r > r_{\Phi}$.  Therein the unit vector in the angular
direction $\hat{\phi}=-\sin(\phi)\,\hat{x}+\cos(\phi)\,\hat{y}$ has
been used. The effect of the linearly polarized {EM} wave is simply
expressed as $\mathcal{\vec{A}}_{EM}(t) = {A_0}\sin{(kz -\Omega t)}
\,\hat{x}$, where $k$ and $\Omega$ are the wave vector and the
frequency of the wave.

The Hamiltonian can be divided as $H = H_0 + H_{\rm int}$, where
$H_0$ and $H_{\rm int}$ correspond to the unperturbed and
time-dependent Hamiltonian respectively. From energy conservation,
the rapidly oscillating quadratic term in
$\mathcal{\vec{A}}_{EM}(t)$ is omitted, and hence
\begin{equation}
 H_{\rm int} \simeq - \{\frac{e\mathcal{\vec{A}}_{EM}}{m^*c}(\vec{p}-
\frac{e}{c}\vec{A}_{\Phi}) +\frac{\alpha e} {\hbar
c}[\vec{\sigma}\times \mathcal{\vec{A}}_{EM}]_z \}.
\end{equation}
It is convenient to rewrite $H_{\rm int}=H_D+H_B+H_{SO}$ indicating
three different types of interaction. The first term is the electron
dipole interaction, given by
\begin{equation}
H_D = \frac{-e}{m^*c}\mathcal{\vec{A}}_{EM}\cdot\vec{p} \simeq -e
\vec{x}\cdot \mathcal{\vec{E}} ,
\end{equation}
where the time-dependent electric field, $\mathcal{\vec{E}}  =
\mathcal{E}_0\cos({\Omega t})\hat{x}$, is polarized along \textbf{x}
direction.  The second contribution $H_B$ is due to the applied
magnetic flux
\begin{eqnarray}
H_B &=& \frac{e^2}{m^* c^2}\mathcal{\vec{A}}_{EM}\cdot
\vec{A}_{\Phi} \nonumber \\
&=& \frac{e^2 \mathcal{E}_0}{m^* c \Omega} A_{\Phi}(r)
\sin(\phi)\sin(\Omega t).
\end{eqnarray}
The third term, $H_{SO}$, denotes the SO coupling mechanism
\begin{eqnarray}
H_{SO}&=&-\frac{\alpha e}{\hbar c}
\left[\vec{\sigma}\times\mathcal{\vec{A}}_{EM}\right]_z\nonumber \\
&=& \frac{-\alpha e \mathcal{E}_0}{\hbar \Omega} \sigma_y
\sin(\Omega t).
\end{eqnarray}

 The electron
dynamics can be derived based on the knowledge of eigenfunctions
of $H_0$. At $t = 0$ the normalized two-component wavefunction is
$\Psi = \left(\Psi_\uparrow,\Psi_\downarrow\right)^T$, where
\begin{equation}
\Psi_{\sigma}=\psi_{\sigma}(\vec{r})\otimes\chi_{\sigma},
\label{Psi}
\end{equation}
with $\sigma=\uparrow$ or $\downarrow$ indicating two spin branches.
Since total angular momentum $J_z$ commutes with time-independent
Hamiltonian in the presence of SOI, the spatial wave function can be
expressed of the form
\begin{eqnarray}\label{spinors}
    \psi_\uparrow(\vec{r})&=&\psi_{\,l_\uparrow}\!(r)\,
    e^{i\,l_\uparrow\phi} \nonumber \\
    \psi_\downarrow(\vec{r})&=&\psi_{\,l_\downarrow}\!(r)\,
    e^{i\,l_\downarrow\phi},
\end{eqnarray}
where the orbital angular momenta $l_\uparrow=m_j-1/2$ and
$l_\downarrow=m_j+1/2$ follow the relation $l_\downarrow =
l_\uparrow +1$, with $m_j$ corresponding to the eigenvalue of
$J_z$.
%\begin{eqnarray}
%\left(%
%\begin{array}{c}
%  \psi_\uparrow \\
%  \psi_\downarrow \\
%\end{array}%
%\right) = \left(%
%\begin{array}{c}
%  \psi_j e^{i(m_j-1/2)\phi} \\
%  \psi_j e^{i(m_j+1/2)\phi} \\
%\end{array}%
%\right), \label{spinors}
%\end{eqnarray}
%and we obtain the relation for angular momentum value:
%$l_\downarrow = l_\uparrow +1$.
While dipole interaction does not flip spin directly the spin
flipping is possibly achieved in the presence of the SOI and
therefore we can investigate the spin-dependent charge dynamics.
To characterize this feature, we define the net spin
polarizability
\begin{eqnarray}
\mathcal{P} =
\frac{\langle\Psi|\sigma_z|\Psi\rangle}{\langle\Psi|\Psi\rangle} =
\frac{\langle\Psi_\uparrow|\Psi_\uparrow\rangle -
\langle\Psi_\downarrow|\Psi_\downarrow\rangle}{\langle\Psi|\Psi\rangle}.
\label{NetPol}
\end{eqnarray}
For the case of $|\mathcal{P}| = 1$, this indicates that the
system is totally polarized into spin-$\uparrow$
(spin-$\downarrow$) if $\mathcal{P}=+1$ ($\mathcal{P}=-1$).
Otherwise, the spin polarizability can be generally specified by
the notation $\mathcal{P}_\uparrow$ if $\mathcal{P} > 0$, and
$\mathcal{P}_\downarrow$ if $\mathcal{P} < 0$.

Specifically, we consider an InAs-based double quantum ring
system, of which the quantum structures are appropriate to
investigate some spin-related phenomena.\cite{InAs} Below we have
selected the InAs material parameters $m^*/m_0 = 0.042$ and
$\alpha \sim 40$ meV$\cdot$nm. Correspondingly, the characteristic
energy $E = \hbar \omega_0 = 5$ meV and for the EM field
$\hbar\Omega = 1$ meV. Dimensionless parameters of the double-ring
potential are: $r_c = 0$, $V_1 = 70$, $V_2 = 20$, $V_3 = -20$,
$\sigma_1 = 1.825$, $\sigma_2 = 1.0$, $\sigma_3 = 2.236$, $r_1 =
0$, $r_2 = 5.0$, and $r_3 = 6.0$.  In the numerical calculation,
we shall present the magnetic flux in units of the flux quantum
$\Phi_0=hc/e$.

\begin{center}
\begin{figure}[thbp!]
\includegraphics[width=0.40\textwidth,angle=0]{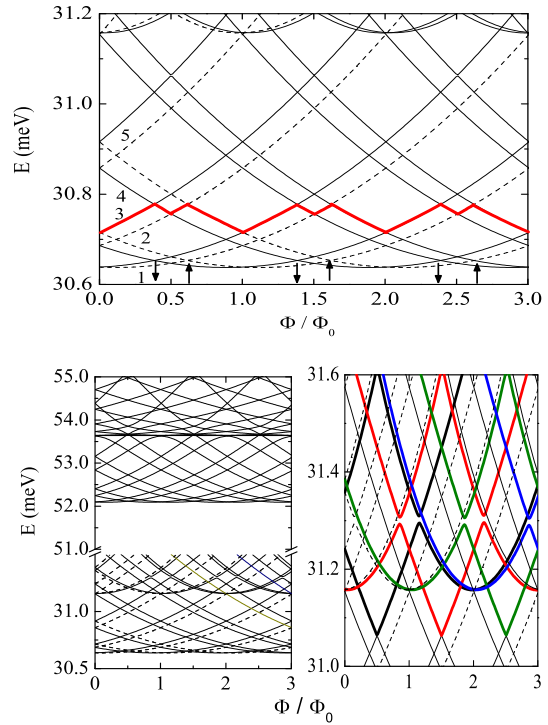}
\caption{\label{L0} (Color online) (a) The Aharonov-Bohm
oscillations in the energy spectrum of the double ring in the
presence of the Rashba SOI with $\alpha=40$ meV$\cdot$\,nm and a
static magnetic flux. The spectrum of states with positive
(negative) $m_j$ are depicted in solid (dashed) curves. The lowest
five pairs of the states are specified by $|m_j| = 2.5,\, 3.5,\,
1.5,\, 4.5$, and 0.5. In each $1/2-\Phi/\Phi_0$ region, up or down
arrows denote net spin orientations of ground states. The zigzag
curve shows the sixth lowest eigenenergy. (b) Energy levels within
lowest four subbands, in which energies of subband bottoms line
close by 30.6, 31.2, 52.1, and 53.6\,meV, respectively. (c) Near the
second subband bottom, anti-crossing levels are depicted in thick
curves. From left to right, these states are of $m_j$ = 0.5 (black),
2.5 (green), 1.5 (red),
  and 3.5 (blue).}
\end{figure}
\end{center}

By choosing the above typical physical quantities and
diagonalizing the time-independent Hamiltonian, it is easy to
obtain the single-particle energy spectrum as is shown in
Fig.~(\ref{L0}a). As compared with the energy spectra in usual
quantum dots, a genuine effect of SOI can be revealed in the
reduction of degeneracies from four-fold to at most two-fold. But
for rings there are only two-fold degeneracies at zero magnetic
flux whether the SOI exists or not. The presence of the magnetic
flux breaks the time reversal symmetry but Kramer's degeneracy is
not lifted due to the absence of the Zeeman effect. In
Fig.~(\ref{L0}a), the solid curves indicate energy levels of
positive $m_j$ whereas those of negative $m_j$ are depicted by
dashed curves. The lowest five pairs of energy levels belong to
$|m_j| = 2.5,\, 3.5,\, 1.5,\, 4.5$, and 0.5. The ordering of these
levels could change if the coupling constant $\alpha$ of the
Rashba SOI is varied. For instance, when $\alpha=5$\,meV$\cdot$nm,
the lowest five pairs of these levels are with $|m_j| = 0.5,\,
1.5,\, 0.5,\, 2.5$, and $1.5$. In the concern of varying $\alpha$,
one can drive coherent ROs and the idea has been realized in
quantum dot systems.\cite{Debald}

In the absence of the magnetic flux, the second and the third (as
well as the fourth and the fifth, etc.) levels in Fig.~(\ref{L0}a)
are close to each other. The gap between these adjacent levels
arises from the zero-field splitting of the Rashba SOI, which will
disappear when the SOI coupling constant $\alpha$ approaches zero.
In the limit $\alpha\rightarrow 0$ not only the adjacent levels at
$\Phi/\Phi_0=0$ mentioned above but also the curves split from
these levels in the region $\Phi/\Phi_0>0$ will merge together.
Another decisive feature distinguishing quantum rings from quantum
dots stands out that for the former the ground state will
periodically shift to that of higher total angular momentum.
However, it always corresponds to the state with the lowest
angular momentum in quantum dots.

An energy level $E$ in Fig.~(\ref{L0}a) is a piecewise smooth
function of $\Phi/\Phi_0$ with singular crossing points. The zigzag
thick curve shows an example of the sixth lowest level, which has
five crossing points around $\Phi/\Phi_0=0.5,\,0.6,\,1.0,\,1.4$ and
$1.5$ within the unit interval $0.5<\Phi/\Phi_0<1.5$ at $\alpha=40$
meV$\cdot$nm. If $\alpha\rightarrow 0$, the pairs of curves merge as
discussed above and the set of crossing points reduce to
$\Phi/\Phi_0=0.5$ and $1.5$ for the ground state, and reduce to
$\Phi/\Phi_0=0.5,\,1.0$, and $1.5$ for other levels. It turns out
that at $\alpha=0$ the electron in the double-ring reveals a similar
oscillation pattern, with the same oscillation period one, as the AB
oscillations in a single ring without SOI. \cite{AB,Chak1994} For
$\alpha > 0$, the splitting of local spin branches in AB oscillating
spectrum of a single ring can be identified.\cite{Governale} For the
double ring, spectral patterns become more complicated, but patterns
with regular oscillations are still rather apparent. Moreover, the
spin polarizability $\mathcal{P}$ defined in Eq.~(\ref{NetPol})
varies in $\Phi/\Phi_0$. It seems that the crossing points of the
lowest level in Fig.~(\ref{L0}a), i.e., at
$\Phi/\Phi_0=\{0.5,\,1,\,1.5,\,2,\,...\}$ are exactly the positions
where $\mathcal{P}$ changes its sign, at least in the range we
studied.

To explain the location of the crossing points in
Fig.~(\ref{L0}a), let us consider an ideal one-dimensional ring of
radius $R$ enclosing a magnetic flux. The spectral property of
this system reflects some key features of a radial subband of the
double ring (Fig.~(\ref{L0}b)). The flux-dependent energy spectrum
can be derived in the analytical form
\begin{widetext}
\begin{equation}
E=\frac{\hbar\omega_a}{2}
\left\{\left(l_\uparrow-\frac{\Phi}{\Phi_0}\right)^2 \!+\!
\left(l_\downarrow-\frac{\Phi}{\Phi_0}\right)^2
\pm\sqrt{\left[\left(l_\uparrow-\frac{\Phi}{\Phi_0}\right)^2
-\left(l_\downarrow-\frac{\Phi}{\Phi_0}\right)^2\right]^2
+\frac{4\alpha^2}{R^2}\frac{1}{\hbar^2\omega^2_a}\left(l_\uparrow-\frac{\Phi}{
\Phi_0}\right)\!\! \left(l_\downarrow-\frac{\Phi}{
\Phi_0}\right)}\right\}, \label{El}
\end{equation}
\end{widetext}
 where $\hbar\omega_a = \hbar^2/2m^*R^2$. Due to the
relation $l_\downarrow=l_\uparrow+1$, the energy $E$ can be
expressed as a function of the variables $l_\downarrow$ and
$\Phi/\Phi_0$. A crossing point will come up at a certain
$\Phi/\Phi_0$ where different integers $l_\downarrow$ have the same
energy $E$. According to Eq.~(\ref{El}), this happens when $\Phi$
increases from zero to a period $\Phi _0$ if the Rashba SOI is
absent. In the presence of the Rashba SOI, additional crossing
points appear before $\Phi$ reaches $\Phi_0$. As compared to the
Fock-Darwin spectrum in a quantum dot where the energy is in linear
proportion to the trapping frequency and the cyclotron frequency in
weak and strong magnetic fields respectively. However, while SOI can
be regarded as the perturbation, Eq.~(\ref{El}) well tells a
quadratic relation with the magnetic flux.

Level crossing points can be easily seen in Fig.~(\ref{L0}a).
Anti-crossing levels also appear, for instance in the example of
Fig.(\ref{L0}c), and even in high energy regimes, as seen in the two
dashed lines at $(\Phi/\Phi_0\,,E)=(1.7,\,53.7)$ in
Fig.~(\ref{L3}a). While the splitting of the accidental level
degeneracy in quantum dots has been demonstrated both theoretically
and experimentally,\cite{split} the repulsions in the avoiding
levels due to the interplay between Zeeman and Rashba terms are also
reported recently.\cite{dot_avoid} In our case, anti-crossings near
the second subband bottom arises in the presence of strong SOI. But
for high energy pairs, the repulsions here are mainly attributed to
the geometric effect of the double ring under influences of the
magnetic flux. In other words, the repulsion levels in the double
ring will not disappear, even when Rashba effect is turned off. In
the vicinity of the minimal splitting points, wavefunctions vary
acutely and cannot be specified by a set of good quantum numbers.
The double-ring Hamiltonian thus typically manifests the signature
of quantum chaos. A comparison between the spectra of $\alpha=0$ and
$20$\,meV$\cdot$nm shows that the Rashba SOI will increase the level
splitting in each energy pair mentioned above but decrease the gap
of the repulsion levels from 0.32 to 0.28\,meV. Moreover, by
adiabatically modulating the gate voltage to change SOI, the
double-ring system discussed here serves as a candidate for testing
the Berry phase.\cite{BerryPhase}

\section{Rabi Oscillations in two-level transitions}

Since an electron in the inner (outer) ring has a definite angular
momentum, its tunneling probability to the neighbor ring is
suppressed under the constrain of the angular momentum
conservation. Therefore, an eigenfunction in the double ring may
be localized only in the inner or the outer ring, if its
corresponding energy is lower than the barrier. In the following
we are going to investigate the dynamics of an electron under
irradiations of an external EM field in the presence of SOI. The
transition between two such kind of quantized levels in the energy
space corresponds to an inter-ring transition in the spatial
space.

\begin{figure}[thbp!]
\includegraphics[width=0.40\textwidth,angle=0]{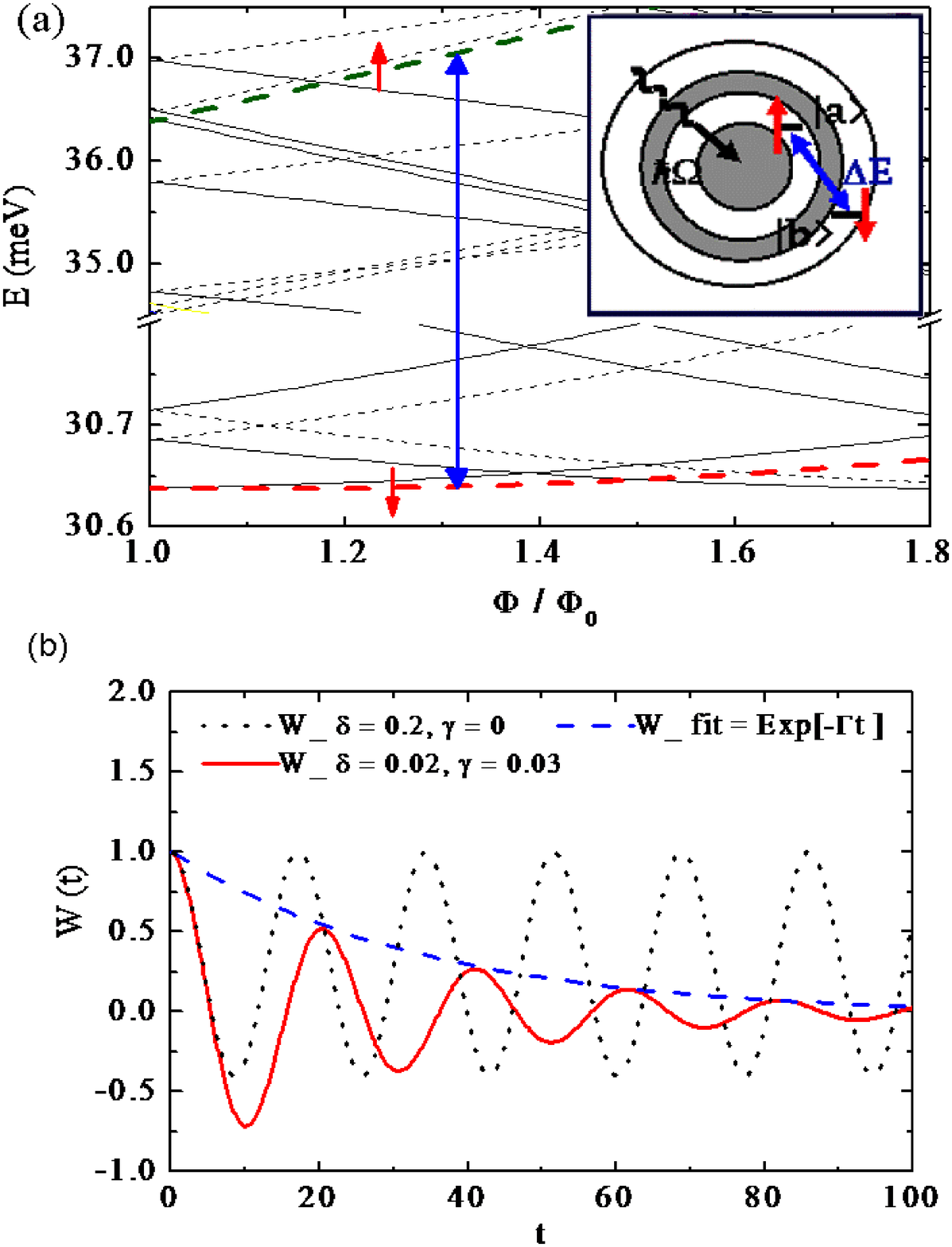}
\caption{\label{L2} (Color online) (a) The energy spectrum
including two states:
$|a\rangle=|-2.5\rangle_{\mathcal{P}_\uparrow}$ and
$|b\rangle=|-1.5\rangle_{\mathcal{P}_\downarrow}$. The sketch in
the inset depicts inter-ring transitions under EM wave
stimulations. (b) Population inversion as the function of time.
When ignoring the spontaneous emission, $W(t)$ with large detuning
is demonstrated by dotted symbol. If the spontaneous emission of
an excited state is considered, then in small detuning regime,
where we set $\gamma =0.03$ and $\delta=0.02$, $W(t)$ manifests an
underdamped oscillation as depicted by the solid curve and the
decay behavior is fitted by an exponential decay function as is
shown by the dashed line.}
\end{figure}

We consider arbitrary two levels in the double ring. Suppose the
electron initially occupies state $|b\rangle$ with eigenfunction
$u_b(\vec{r})$ in the outer ring, the irradiation process is
designed to pump it to state $|a\rangle$ with eigenfunction
$u_a(\vec{r})$ in the inner ring. For convenience the
time-dependent wave function can be written as
\begin{equation}
  {\bf\psi}(\vec{r},t) = c_a(t)e^{i(\delta/2-\omega_a)t} u_a(\vec{r})+ c_b(t)e^{i(-\delta/2-\omega_b) t}
 u_b(\vec{r}), \label{Eq_2state}
 \end{equation}
where $u_a(\vec{r})$ and $u_b(\vec{r})$ associated with $E_{j} =
\hbar\omega_{j}$ for $j=a$ and $b$, related to two-component wave
function $\Psi$ in Eq.~(\ref{Psi}), as eigenfunctions and
eigenenergies of $H_0$. Moreover, an optical transition takes
place between two states that correspond to dipole-allowed
eigenstates conforming to the relation $\Delta m_j = \pm 1$. As
usual, $\delta \equiv (\Delta E/\hbar - \Omega)$ is the detuning
defined as the frequency difference between the level spacing and
the laser field as shown in the inset of Fig.~(\ref{L2}a). Insert
Eq.~(\ref{Eq_2state}) into time-dependent Schr\"{o}dinger
equation, the time-evolution of an electron can be expressed as
\begin{eqnarray}
\left[%
\begin{array}{c}
  \dot{c_a}(t) \\
  \dot{c_b}(t) \\
\end{array}%
\right] &=& \frac{i}{2} \left[%
\begin{array}{cc}
 -\delta & R_{D}+ \tilde{R} \\
  R_{D}^*+ \tilde{R}^* & \delta \\
\end{array}%
\right]
\left[%
\begin{array}{c}
  c_a(t) \\
  c_b(t) \\
\end{array}%
\right], \label{Dynamics2}
\end{eqnarray}
with $\tilde{R} = R_{A}+R_{SO}$, which after some calculations
have the relations
\begin{eqnarray*}
   R_D &=& e \mathcal{E}_0 \langle u_a({r})|{r}|u_b({r})\rangle/4\hbar, \\
   R_B &=& e^2 \mathcal{E}_0 \langle
       u_a({r})|A_{\Phi}(r)|u_b({r})\rangle/4\hbar m^*c\Omega, \\
   R_{SO} &=& \alpha e \mathcal{E}_0 \langle
      u_a({r})|u_b({r})\rangle/2\hbar^2
      \Omega.
\end{eqnarray*}
Therein, $R_D$ is the dipole-induced Rabi frequency as usually
discussed and $R_B$ and $R_{SO}$ denote the couplings of the laser
field with the magnetic flux respectively the Rashba SOI. In
calculating $R_{SO}$, only inner products between partial waves
with different spin orientations would be taken into account. In
deriving Eq.~(\ref{Dynamics2}), we utilized the rotating-wave
approximation (RWA) and ignored the counter-rotating terms
proportional to $\exp[\pm i(\Omega+ \Delta E/\hbar)t]$.

For an electron initially occupies the low energy state
$|b\rangle$ the time-dependent population probability can be
exactly written as
\begin{eqnarray}
|c_a|^2 &=& \frac{\tilde{R}_{\rm eff}^2}{\Omega_d^2}\,\sin^2\!\left(\frac{\Omega_d t }{2}\right) \nonumber\\
|c_b|^2 &=&1- \frac{\tilde{R}_{\rm
eff}^2}{\Omega_d^2}\,\sin^2\!\left(\frac{\Omega_d t }{2}\right),
\label{eqn2}
\end{eqnarray}
where $\Omega_d^2 = \tilde{R}_{\rm eff}^2 +\delta^2 $ and
$\tilde{R}_{\rm eff} = R_D +\tilde{R}$ can be regarded as the
\textit{effective Rabi frequency} in the presence of the external
fields and the SOI. The on-resonance transitions occur when $\delta
= 0$, in the case the population probability can be simply reduced
to $|c_a|^2 =\sin^2 ({\tilde{R}_{\rm eff}t/2}) $ and $ |c_b|^2 =
\cos^2 ({\tilde{R}_{\rm eff}t/2}) $.

If the energy dissipation to the environment is considered as the
interaction between an electron with continuous vacuum modes, a
phenomenological decay parameter $\gamma$ will be introduced to the
first element of the matrix in Eq. (\ref{Dynamics2}), which opens a
decay path from the excited state to its surrounding. Thus the
coefficients describing the system will be changed in the more
complicated form
\begin{widetext}
\begin{eqnarray}
|c_a|^2 &=& \frac{\tilde{R}_{\rm eff}^2}{\tilde{\Omega}^2}e^{-\gamma
t}e^{-\tilde{\Omega} \zeta t}
\sin^2\!\left({\frac{\tilde{\Omega}\eta t}{2}}\right)\nonumber\\
|c_b|^2 &=& \frac{e^{-\gamma t}}{\tilde{\Omega}^2}
e^{-\tilde{\Omega} \zeta t}
\left[(\gamma^2+\delta^2)\sin^2\!\left({\frac{\tilde{\Omega}\eta
t}{2}}\right)
+\tilde{\Omega}^2\cos^2\!\left({\frac{\tilde{\Omega}\eta
t}{2}}\right) +
2\tilde{\Omega}(\gamma\eta-|\delta|\zeta)\sin\left({\tilde{\Omega}\eta
t}\right)\right],
\end{eqnarray}
\end{widetext}
where $\zeta=\cos({{\theta_1}/{2}})$ and
$\eta=\sin({{\theta_2}/{2}})$, with
\begin{eqnarray*}
  \theta_1 &=& \cos^{-1}[{(\gamma^2-\Omega_{d}^2)/\tilde{\Omega}^2}] \\
  \theta_2 &=& \sin^{-1}[{2|\delta|\gamma/\tilde{\Omega}^2}] \\
  \tilde{\Omega}^2 &=& \sqrt{(\gamma^2-\Omega_{d}^2)^2
+(2\delta\gamma)^2 },
\end{eqnarray*}
in which $\theta_1$ and $\theta_2$ should be taken from the same
quadrant.

For convenience, we introduce the notation
$|m_j\rangle_{\mathcal{P}}$ to specify a state in terms of the
total angular momentum $m_j$ and its spin polarizability
$\mathcal{P}$. We choose two states $|a\rangle=
|-2.5\rangle_{\mathcal{P}_\uparrow}$ and
$|b\rangle=|-1.5\rangle_{\mathcal{P}_\downarrow}$ respectively, at
$\Phi=1.3\,\Phi_0$ with $E_a = 37.03$ meV and $E_b = 30.64$ meV,
as depicted in Fig.~(\ref{L2}a). The corresponding population
inversion
\begin{equation}
W(t)=|c_b(t)|^2-|c_a(t)|^2
\end{equation}
of this system is demonstrated in Fig.~\ref{L2}(b). If the vacuum
fluctuation is absent ($\gamma=0$), the total probability is
conserved, i.e., $|c_b(t)|^2+|c_a(t)|^2 = 1$ at any time. In this
case $W(t)$ manifests oscillating behavior within the interval
$[1-2(\tilde{R}_{\rm eff}^2/\Omega_d^2),1]$ without any
dissipation. For $\delta\rightarrow 0$, this interval approaches
to its maximum values $[-1,1]$, namely that $\Omega_d\rightarrow
\tilde{R}_{\rm eff}$. For large detuning, e.g. $\delta=0.2$, this
interval will shrink to its $70\%$, and the blue shifted inversion
curve is plotted in the dotted line in Fig.~(\ref{L2}b).

If the spontaneous emission of an excited state is considered,
$W(t)$ will decay with time, as shown by the solid curve in
Fig.~(\ref{L2}b), in which $\gamma =0.03$ and $\delta=0.02$. In
this case $W(t)$ oscillates underdamped. The damping behavior
which in consistent with the Weisskopf-Wigner theory\cite{Weiss}
is well-fitted by $W_{\rm fit} = e^{-\Gamma t}$, where the Fermi's
rate $\Gamma \sim 0.03$. When the Rabi relaxation time is defined
as $\tau_{R}= 1/\Gamma$, our calculation shows that a cycle of
inter-ring transition accomplishes in $\tau_R$ for a given
$\gamma$.
%"Spontaneous decay rate is proportional to the density of state."
While the spontaneous decay can be experimentally controlled and
suppressed,\cite{Spondecay} in a cavity with a limited number of
modes at the transition frequency, a long decoherence time is
permitted.
%within
%electron-photon interactions can be expected.
In assumption of weak system-environment coupling, efficient
population transfers are feasible. So in conclusion, under SOI we
can simultaneously manipulate electron transitions associated with
its spin orientations in either rings via the RO processes of a
two-level model.

\section{The photon-assisted transitions in three-level schemes}

\begin{figure}[b!]
\includegraphics[width=0.35\textwidth,angle=0]{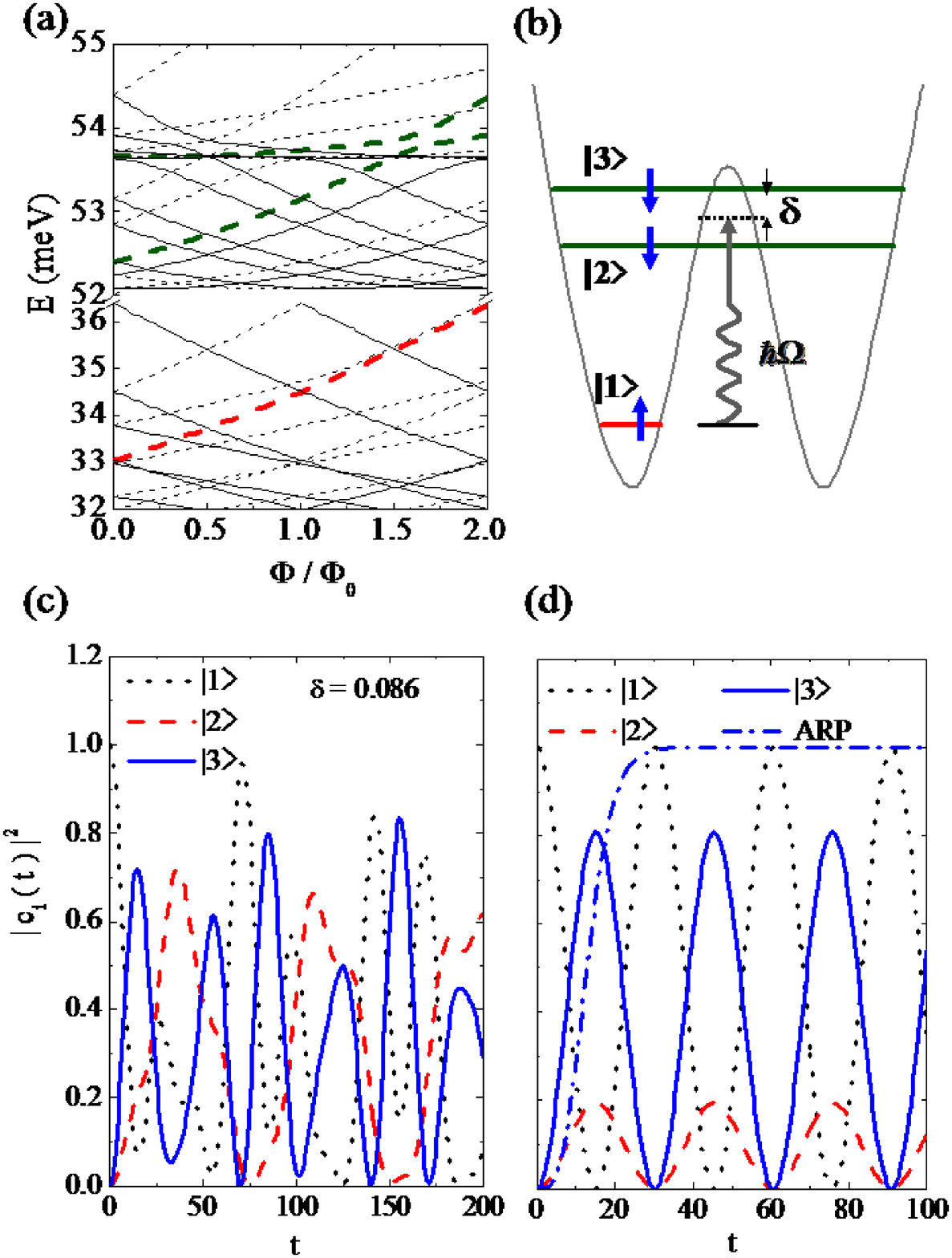}
\caption{\label{L3} (Color online) (a) The energy spectrum
including the chosen three levels, an up-spin state
$|1\rangle=|-1.5\rangle_{\mathcal{P}_\uparrow}$ (red), and a
down-spin doublet
$|2\rangle=|-2.5\rangle_{\mathcal{P}_\downarrow}$ and
$|3\rangle=|-2.5\rangle_{\mathcal{P}_\downarrow}$ (green) at
$\Phi= 1.7 \Phi_0$. Level repulsion occurs between the doublet,
and the gap is about 0.17\,meV for $\alpha=40$\,meV$\cdot$nm. (b)
A sketch of a cascade-type model in the double ring. population as
a function of time for the cases of (c) $\delta=0.086$ and (d)
$\delta\rightarrow 0$. Dash-dotted curve in (d) shows the ARP
estimation of the transition from $|2\rangle$ to $|3\rangle$.}
\end{figure}

In this section we apply the Rabi model to several interesting
cases. We show how inter-ring transition are achieved via the
photon-assisted processes. The processes are demonstrated by
considering three-level systems in cascade-type and $\Lambda$-type
schemes that are shown in Fig.~(\ref{L3}b) and Fig.~(\ref{L4}b).
For clarity, we rewrite the time-dependent wave function as
$\psi(\vec{r},t) = \sum_{j=1}^3 c_j(t)e^{-i\omega_j t}
u_j(\vec{r})$, where $E_j =\hbar\omega_j$ and we define
$E_{12}=E_2-E_1$, $E_{13}=E_3-E_1$, $E_{12} + \delta = E_{13} -
\delta = \hbar\Omega$. The involved states construct two
dipole-allowed transitions $\{|1\rangle \leftrightarrow
|2\rangle\}$, $\{|1\rangle \leftrightarrow |3\rangle\}$ and a
$\{|2\rangle \leftrightarrow |3\rangle\}$ dipole-forbidden paths.
For a clear demonstration we shall ignore spontaneous emission
processes.

\subsection{The cascade-type}

It is well known that the quantum-beat spectroscopic method
permits the resolution of closely neighboring levels.\cite{beat}
Earlier experiments demonstrate that quantum-beat spectroscopy is
a useful technique in the measurement of Zeeman splittings and
hyperfine intervals in atomic and molecular
systems.\cite{beatmeasure} It is then interesting to study
spectroscopic dynamics involving the \textit{direct} inter-ring
transitions in the cascade-type scheme nearby the avoided crossing
points.

Again, we use the notation $|m_j\rangle_{\mathcal{P}}$ to specify
the involving states. Here we consider an up-spin state
$|1\rangle=|-1.5\rangle_{\mathcal{P}_\uparrow}$, and a down-spin
doublet $|2\rangle=|-2.5\rangle_{\mathcal{P}_\downarrow}$ and
$|3\rangle=|-2.5\rangle_{\mathcal{P}_\downarrow}$ at $\Phi= 1.7
\Phi_0$. For state $|1\rangle$, the electron is localized in the
inner ring. The electron wave functions, however, extend over the
double ring for two higher states. The energy spectrum is depicted
in Fig.~(\ref{L3}a), and in (b) we show the sketch of the cascade
model. Similar to solving Eq.~(\ref{Dynamics2}) for the two-level
system, transitions among different spin states can be
investigated through
%\begin{widetext}
\begin{eqnarray}
\dot{c_1}(t)&=&\frac{i}{2}\left[e^{-i\delta t}R^{12}_{+}c_2(t)+
e^{i\delta t}R^{13}_{-}
 c_3(t)\right]\nonumber\\
\dot{c_2}(t)&=&\frac{i}{2}e^{i\delta
t}R^{*21}_{+}c_1(t) \nonumber\\
\dot{c_3}(t)&=&\frac{i}{2} e^{-i\delta t}R^{*31}_{-}c_1(t),
\label{Eq3state}
\end{eqnarray}
where $R^{ij}_{\pm} = R_D^{ij}\pm \tilde{R}^{ij}$. Adopting
transformations $c_2(t)=C_2(t)e^{i\delta t}$ and
$c_3(t)=C_3(t)e^{-i\delta t}$, Eq.~(\ref{Eq3state}) can be
reexpressed as a homogeneous autonomous equation. For an electron
initially occupies state $|1\rangle$, the population probability
in small detuning regime can be expressed approximately to
\begin{eqnarray}
|c_1(t)|^2 &=& \cos^2 \left(\frac{M_{R}t}{2}\right) \nonumber\\
|c_2(t)|^2 &=& \left(\frac{{R^{12}_{+}}}{M_{R}}\right)^2 \sin^2 \left(\frac{M_{R}t}{2}\right)\nonumber\\
|c_3(t)|^2 &=& \left(\frac{{R^{13}_{-}}}{M_{R}}\right)^2 \sin^2
\left(\frac{M_{R}t}{2}\right), \label{3cas}
\end{eqnarray}
where $M_R = \sqrt{(R^{12}_{+})^2+(R^{13}_{-})^2}$. Beyond the
small detuning approximation, Eq.~(\ref{Eq3state}) is numerically
solved and the result for $\delta=0.086$ is shown in
Fig.~(\ref{L3}c). Since transition probabilities come up
differently between two paths, namely $R^{12}_{+}\neq R^{13}_{-}$,
occupations with time on each states turn into be aperiodic and
less regular compared with probabilities obtained from
Eq.~(\ref{3cas}). Discrepancies between Fig.~(\ref{L3}c) and
Fig.~(\ref{L3}d) are clearly illustrated. However, in the case
$\delta\rightarrow 0$, the electron tends to oscillate between
$|3\rangle$ and $|1\rangle$ and the transition
$|2\rangle\leftrightarrow|3\rangle$ is less efficient, as compared
with the off-resonance transitions.

This drawback can be removed by another interesting manipulation in
the cascade scheme, namely, by transferring electrons ladder by
ladder with chirped laser pulses. The idea originates from the
electron transfer in molecules, in which stepwise excitations are
applied for a rapid and efficient dissociation of specific chemical
bonds.\cite{ladder} Under ARP condition, electrons that are
resonantly pumped to state $|2\rangle$ could be efficiently
transited to the final state $|3\rangle$ in a long trapping time.

To this end, we select a chirped laser pulse with time-dependent
electric field along the radial direction, given by
\begin{equation}
\vec{\mathcal{E}}_{ch}(t)= \mathcal{E}_{ch}
\exp\left(-\frac{t^2}{2\tau^2}-i\Omega_{ch} t -
i\beta\frac{t^2}{2}\right)\hat{r}, \label{Et}
\end{equation}
%is chosen to drive stimulated transitions between states
%$|2\rangle$ and $|3\rangle$,
where $\tau$ is the pulse duration, $\Omega_{ch}$ is the central
frequency, and $\beta$ is the temporal chirp. The equation of
motion in RWA modified from Eq.~(\ref{Dynamics2}) becomes
\begin{eqnarray}
\left[%
\begin{array}{c}
  \dot{c_2}(t) \\
  \dot{c_3}(t) \\
\end{array}%
\right] &=& \frac{i}{2} \left[%
\begin{array}{cc}
  0 & R_{ch}(t) \\
  R_{ch}^*(t) & 2\delta_{ch}(t) \\
\end{array}%
\right]
\left[%
\begin{array}{c}
  c_2(t) \\
  c_3(t) \\
\end{array}%
\right],
\end{eqnarray}
in which $\delta_{ch}(t) = \beta t$ is linearly chirped detuning,
and $R_{ch}(t)$ stands for the time-dependent Rabi frequency
related the pulse envelope of $\vec{\mathcal{E}}_{ch}(t)$. By
choosing proper parameters for a pulse with peak Rabi frequency
$R_{0} = 0.25$, $\beta = 0.01$, and $\tau = 10$,
%If electrons can
%be resonantly pumped from $|1\rangle$ to $|2\rangle$, they will
%tend to move to and stay at $|3\rangle$ stably (the dash-dotted
%line in Fig.~\ref{L3}(d)).
a complete transfer is demonstrated by the dash-dotted line in
Fig.~\ref{L3}(d)). The numerical result coincides with the
estimation of the Landau-Zener formula \cite{LZ}
\begin{equation}
P \simeq 1- \exp\left(-\pi \frac{R_{0}^2}{2\beta}\right).
\end{equation}
In the adiabatic limit
%\begin{equation}
$|\beta| \ll R_{0}^2$,
%\end{equation}
electrons have great probability to occupy an excited state in the
long time limit. Distinguished time-evolution populations between
stimulated transitions by CWs and steady transfer by a chirp pulse
are depicted by curves in Fig.~(\ref{L3}d). Therefore, in addition
to the manipulation of electron transitions between a single ring
and a double ring via the subjection to CW irradiations, we also
arrive at the optimal control on stably selective excitations with
chirped pulses in cascade-type systems.

\subsection{The $\Lambda-$type transition}

Finally we shall also investigate an interesting and important
phenomenon: Whenever there are level crossings for two energy
states that belong to either one ring, $\Lambda$-type scheme of
indirect inter-ring transitions among which and one higher energy
state can be switched on. Applications of this model has been
proposed both in superconducting quantum interference device
(SQUID)\cite{Chu} and semiconductor double quantum
dots,\cite{mlevelqubit} in which multi-level ROs as a target
towards coherent control has been demonstrated.

\begin{figure}[pbt!]
\includegraphics[width=0.35\textwidth,angle=0]{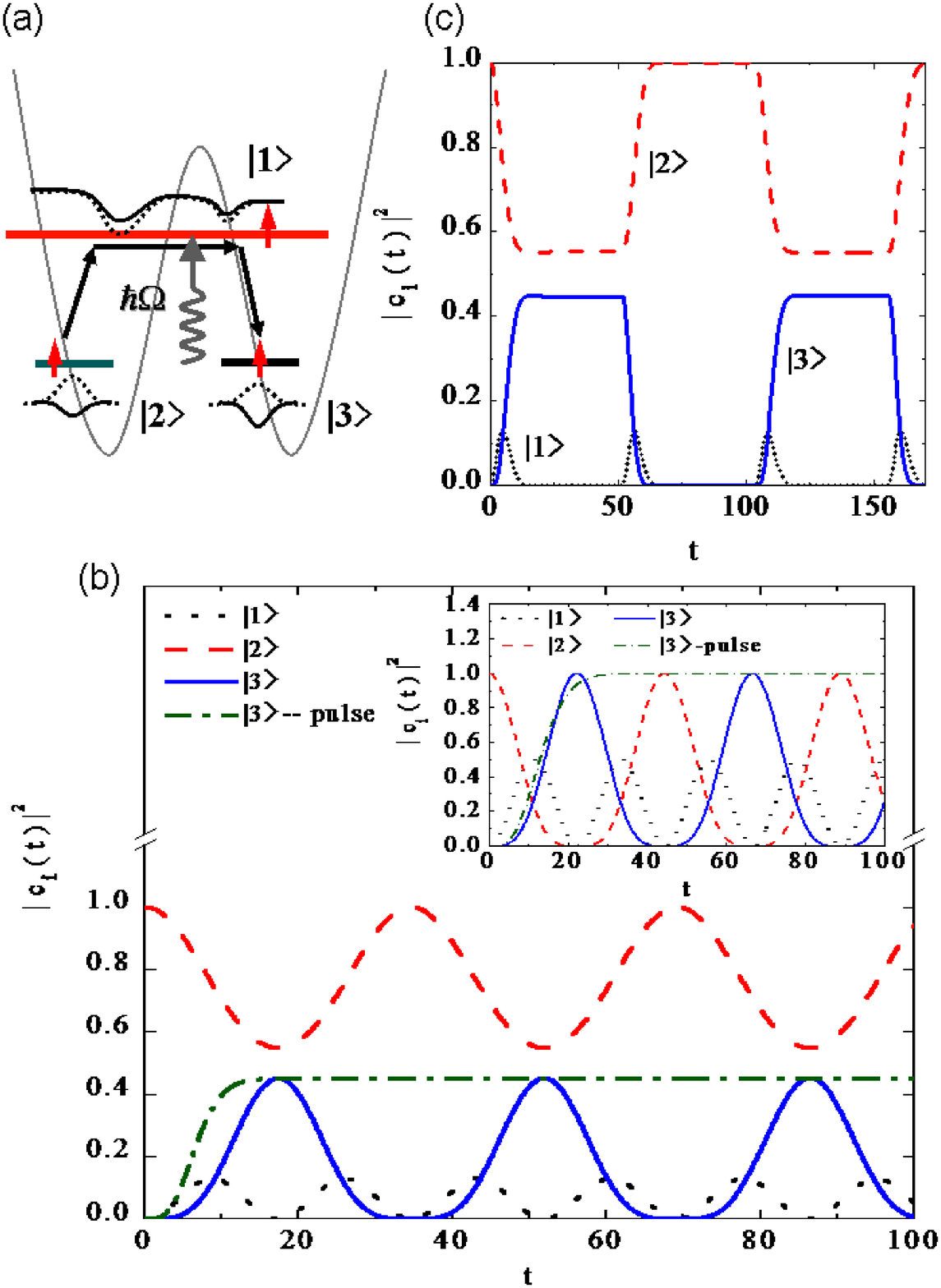}
\caption{\label{L4} (Color online) (a) A sketch of the
$\Lambda-$type scheme. States here are all with effective up-spin
orientations. An electron initially occupies state $|2\rangle$ can
be optically pumped to an outer-ring state $|3\rangle$ mediated by
state $|1\rangle$. (b) The population probabilities among three
levels. The long-time occupation of an excited state is feasible by
applying an short and intense pulse. In the inset we show the
optimal transfer is feasible provided that there is common Rabi
frequency in the two paths. (c) An alternative output signal can be
obtained under successive pulse stimulations.}
\end{figure}

As Fig.~(\ref{L4}a) shows that for an electron occupies state
$|2\rangle$ the photon-assisted quantum transition is initiated
from the inter-ring surpassing an intermediate-state $|1\rangle$
and finally reach to the outer-ring of state $|3\rangle$.
On-resonance solutions of the
\textit{mediated-indirect-transition} system with initial
conditions $c_1(0)=0, c_2(0)=1$, and $c_3(0)=0$ are
\begin{eqnarray}
|c_1(t)|^2 &=& \left(\frac{R^{12}_{+}}{M_{R}}\right)^2 \sin^2\left(\frac{M_{R}t}{2}\right)\nonumber\\
|c_2(t)|^2 &=&
\left(\frac{R^{12}_{+}}{M_{R}}\right)^4\cos^2\left(\frac{M_{R}t}{2}\right)+\left(\frac{R_{-}^{13}}
               {M_{R}}\right)^4\nonumber\\
         &&+\frac{2(R_{+}^{12} R_{-}^{13})^2}{M_{R}^4}\cos\left(\frac{M_{R}t}{2}\right) \nonumber\\
|c_3(t)|^2 &=&
\frac{\left(R_{+}^{12}R_{-}^{13}\right)^2}{M_{R}^4}\left[\cos\left(\frac{M_{R}t}{2}\right)
-1\right]^2, \label{Lambdaeqn}
\end{eqnarray}
and the transition probabilities are shown in Fig.~(\ref{L4}b).
The incomplete transfer and occupation are restricted due to
unequal effective Rabi frequencies in two paths. Once the external
field optimizes the efficiency of one path, the efficiency of the
other is not optimal.

The formulae in Eq.~(\ref{Lambdaeqn}) clearly show that the
probability $|c_3(t)|^2$ has a maximum value at $t=m\pi/M_R$ with
an odd integer $m$ and a minimum value at $t = n\pi/M_R$ with an
even integer $n$. At these extreme points we have the ratio
$|c_2(t)|^2/|c_3(t)|^2 =
\left[(R_{+}^{12})^2-(R_{-}^{13})^2\right]^2/4(R_{+}^{12})^2(R_{-}^{13})^2$,
which indicates the optimal transfer at
$|R_{+}^{12}|/|R_{-}^{13}|=1$. Away from this ratio, the transfer
efficiency will decrease. In Fig.~(\ref{L4}b), transition among a
doublet of $m_j = 0.5$ and 2.5 and an auxiliary level of $m_j=1.5$
at $\Phi=0.8\, \Phi_0$ is considered. In this case, there is only
45$\%$ occupation on $|3\rangle$ for $R_{+}^{12}/R_{-}^{13}\sim$
0.38. If the effective Rabi frequencies of the paths
$2\leftrightarrow 1$ and $1 \leftrightarrow 3$ are the same, the
optimal control of electron dynamics is feasible. To show this, we
calculate the time-dependent occupation probability on each
states. In the inset, it is clear that at extreme times,
occupation on $|3\rangle$ based on a pseudo $\Lambda-$ transition
process has a maximum value. Meanwhile, states $|1\rangle$ and
$|2\rangle$ are empty left. Moreover, we also investigate the
transition process stimulated by chirped pulse irradiation. By
setting $R_{0}=0.25$, $\beta=0.001$, and $\tau=14.15$, we obtain a
long-time occupation of the final state, obeying estimation of
Landau-Zener's formula. The result is shown by dash-dotted line in
the inset.

Within proper controls of the pulse width $\tau$ against the typical
level spacings and the dephasing time of electrons, it is easy to
manipulate electronic states in $\Lambda-$scheme system by
successive application of pulses.\cite{square} Using the same
parameters as in Fig.~(\ref{L4}b), we simulate the pulse-induced
periodic oscillations. Here the Gaussian pulse duration $\tau=7.3$
and the pulse interval is about $7\tau$. Different from the
sinusoidal ROs, an alternative square wave is shown in
Fig.~(\ref{L4}c). Apparently, the level occupation time in both
inner and outer ring is prolonged. Moreover, since the duration of
the pulse is properly tuned, it can be expected that time evolution
of the occupation should be complete in the pseudo $\Lambda-$
transition process. Otherwise, under-excitation or over-excitation
takes place corresponding to the duration being too short or too
long, respectively. Level occupation will never be or just be
transiently complete. The well control of pulse delay time also
shows the flexibility of manipulations on quantum states.

\section{Conclusion}

In this work we have studied magnetic-optical transitions in a
semiconductor double ring in the presence of Rashba spin-orbit
coupling and the magnetic flux. First, based on accurate numerical
calculations we obtain SOI-accompanied AB energy spectra and
corresponding eigenstates. The presence of the SOI has important
influences on the occurrence of level crossings showing the
evidence both for the periodic orbital motion and spin flips. In
addition, there are anti-crossing levels playing the role of the
magnetic-resonant extraditions of electrons between inner and
outer rings. In high energy regime, occurrence of avoided
crossings indicates a chaotic signature in its classical analogy.
To facilitate the peculiar features of the double-ring system we
have designed some interesting dynamic processes such that the
system can be easily explored experimentally in the near future.

We have studied the temporal evolution processes in a two-level
and two three-level models. The interaction between external
fields and the electron results in the successive stimulating
absorption and emission of a photon and turn out as the effective
Rabi oscillators. In the two-level model, we demonstrate an
alternative manipulation of electrons transiting between two
rings. In cascade scheme, aperiodic and incomplete population
transfer is revealed under the sinusoidal field excitations. By
appropriate tuning SOI strength, the gap between avoided crossing
levels can be reduced such that the rectified output signals are
measurable. Moreover, by short pulse excitations we also
demonstrate the possibility on optimal control of selective and
direct signals. In this work, only one lader transition is
demonstrated, which however can be extended to multi-ladder
transitions following the same principle. Finally, we have
explored the photon-assisted tunneling in the $\Lambda-$type
model. In addition to generation of ROs also we give the criterion
of the most efficient transfer via the mediated-indirect-tunneling
paths. Further, by successive pulse irradiations the well control
on pulse delay results in the time prolongation on state
populations. We should also emphasize that in the $\Lambda-$type
scheme the minimization of the intermediate-level population is
achieved which is an important and practical strategy in device
realization.

The presence of SOI allows the manipulation of spin degree of
freedom and it is timely to examine the spin-dependent optical
response. Since, similar features could be found in double-dot
systems, the above theoretical results in a double ring might shed
light on the future experimental findings in these burgeoning
quantum systems. While there are few works on optically induced and
SOI driven spin dynamics in quantum
systems,\cite{Debald,spindot,PRB193309} we believe that the
theoretical and experimental works of related spin read-out
information by optical pumping in ring-like systems could be carried
out in the near future.

\begin{acknowledgments}
This work was supported by National Science Council and Academia
Sinica in Taiwan. The authors are grateful to valuable discussions
with V. Gudmundsson, Y. N. Chen, G. Y. Chen, and W. Xu.
\end{acknowledgments}

\end{document}